\documentclass[%
reprint,superscriptaddress,amsmath,amssymb,aps,floats,floatfix,showpacs,prd,twocolumn,nofootinbib,nolongbibliography]{revtex4-2}
\usepackage{mathtools,needspace,enumitem,etoolbox,physics,microtype,afterpage,bigints,tabularx,xspace}
\usepackage{graphicx, dcolumn, bm, lipsum}
\usepackage[commentmarkup=footnote]{changes}
\definecolor{linkcolor}{rgb}{0.0,0.3,0.5}
\usepackage[colorlinks = true,
            linkcolor = linkcolor,
            urlcolor  = linkcolor,
            citecolor = linkcolor,
            anchorcolor = linkcolor]{hyperref}  
\usepackage{orcidlink}
\usepackage{soul}
\usepackage{pifont}
\usepackage{aasmacros}

\DeclarePairedDelimiter\floor{\lfloor}{\rfloor}
\usepackage{array}
\definechangesauthor[name={RB}, color=teal]{RB}

\newcommand{\model}[1]{\textsc{M#1}}
\usepackage{acronym}

\begin{document}
\preprint{APS/123-QED}
\title{Test for LISA foreground Gaussianity and stationarity: extreme mass-ratio inspirals}
\newcommand{\bham}{\affiliation{Institute for Gravitational Wave Astronomy \& School of Physics and Astronomy, University of Birmingham, Birmingham, B15 2TT, UK}}
\newcommand{\milan}{\affiliation{Dipartimento di Fisica ``G. Occhialini'', Universit\'a degli Studi di Milano-Bicocca \& INFN, Sezione di Milano-Bicocca, Piazza della Scienza 3, 20126 Milano, Italy}}
\newcommand{\como}{\affiliation{Dipartimento di Scienza e Alta Tecnologia, Universitá degli Studi dell’Insubria, I-22100 Como, Italy}}
\newcommand{\tou}{\affiliation{Laboratoire des 2 Infinis - Toulouse (L2IT-IN2P3), Université de Toulouse, CNRS, UPS, F-31062 Toulouse Cedex 9, France}}

\author{Manuel Piarulli~\orcidlink{0009-0009-4099-9166}}
\email{manuel.piarulli@l2it.in2p3.fr}\tou\milan
\author{Riccardo Buscicchio~\orcidlink{0000-0002-7387-6754}}
\email{riccardo.buscicchio@unimib.it}\milan\bham
\author{Federico Pozzoli~\orcidlink{0009-0009-6265-584Xl}}\como
\author{Ollie Burke~\orcidlink{0000-0003-2393-209X}}\tou
\author{Matteo Bonetti~\orcidlink{0000-0001-7889-6810}}\milan
\author{Alberto Sesana~\orcidlink{0000-0003-4961-1606}}\milan

\date{\today}

\begin{abstract}
Extreme Mass Ratio Inspirals (EMRIs) are key observational targets for the Laser Interferometer Space Antenna (LISA) mission. Unresolvable EMRI signals contribute to the formation of a gravitational wave background (GWB). 
Characterizing the statistical features of the GWB from EMRIs is of great importance, as EMRIs will ubiquitously affect large segments of the inference scheme.
In this work, we apply a frequentist test for GWB Gaussianity and stationarity, exploring three astrophysically-motivated EMRI populations. We construct the resulting signal by combining state-of-the-art EMRI waveforms and a detailed description of the LISA response with time-delay interferometric variables.
Depending on the brightness of the GWB, our analysis demonstrates that the resultant EMRI foregrounds show varying degrees of departure from the usual statistical assumptions that the GWBs are both Gaussian and Stationary.
If the GWB is non-stationary with non-Gaussian features, this will challenge the robustness of Gaussian-likelihood model, when applied to global inference results, e.g. foreground estimation, background detection, and individual-source parameters reconstruction.

\end{abstract}

\maketitle
\section{\label{sec:intro} Introduction}
The Laser Interferometer Space Antenna (LISA) is expected to be the first mission to detect gravitational waves (GWs) from space, operating in the low-frequency band from $0.1 {\rm mHz}$ up to $1 {\rm Hz}$~\cite{2024arXiv240207571C}. 
Unlike terrestrial interferometers, which detect short-lived transient events, multiple overlapping signals will persist in the LISA data stream for its entire duration.
Given the bandwidth of low frequencies it will be sensitive to, it is expected that LISA will observe a wide variety of GW sources~\cite{2017arXiv170200786A, 2024arXiv240207571C}.
In this study, we focus on extreme mass ratio inspirals (EMRIs): binary systems composed of a stellar-mass compact object (CO) and a massive black hole (MBH), with masses of $1-100 M_\odot$ and $10^4-10^7 M_{\odot}$, respectively~\cite{2024arXiv240207571C}.
EMRIs are probes for characterizing COs populations and their dynamics in the host galactic nuclei~\cite{2019BAAS...51c..42B}.
These light COs undergo many orbital cycles before crossing the MBH event horizon~\cite{2015JPhCS.610a2002A}, thus providing an excellent probe of the spacetime geometry around the central compact object. 
This enables stringent tests of the ``no hair'' theorem~\cite{2013LRR....16....7G, 2023PhRvL.130x1402W},  and more broadly of general relativity (GR) in its strong-field regime.
Furthermore, due to the complex dependence of the GW phase on the source parameters, EMRIs provide unparalleled precision measurements of the binary astrophysical parameters~\cite{2017PhRvD..95j3012B}.

The majority of EMRIs are expected to be too faint to be individually detectable~\cite{2017PhRvD..95j3012B, 2020PhRvD.102j3023B, 2023PhRvD.108j3039P}. 
Their GW signals accumulate in an incoherent superposition, forming a stochastic GW background (GWB) in LISA. 
The majority of data analysis algorithms within the literature assume their target GWBs to be stationary and Gaussian~\cite{2020CQGra..37u5017K, 2019JCAP...11..017C,2023JCAP...04..066B,2023PhRvD.107l3531H, 2024PhRvD.109h3029P} (for an extension to the latter, see, e.g. Ref.~\cite{2023PhRvD.108j3005S}).
However, EMRIs exhibit high eccentricities [0.1--0.9] and their GW emission is typically broadband~\cite{1963PhRv..131..435P}, with a resulting non-trivial spectrum.
For this reason, in this work we investigate thoroughly the contribution of individual emissions to the collective GWB, and their effect on its stationarity and Gaussianity.

The paper is organized as follows:
in Sec.~\ref{sec:forestatprop} we introduce the formalism to describe a GWB and its statistical properties; in Sec.~\ref{sec:emri_pop} we construct a series of astrophysically motivated EMRI populations, to capture the large uncertainties in their formations channels~\cite{2017PhRvD..95j3012B}; in Sec.~\ref{sec:gwb}, we generate each source signal using state-of-the-art waveform models~\cite{2021PhRvD.104f4047K} and LISA response, in order to evaluate its detectability; in Sec.~\ref{sec:gwb_eval} we then combine all unresolvable source signals into simulated GWBs. 
In Sec.~\ref{sec:Ray_test} we outline the properties of the statistical test performed on the resulting GWBs, following closely the approach in Ref.~\cite{2024arXivDWD}. 
In Sec.~\ref{sec:discussion} we present the results of our test applied to the simulated EMRI foregrounds. 
Finally, in Sec.~\ref{sec:conclusions} we summarize our findings and discuss future improvements.

\section{Foreground statistical Properties\label{sec:forestatprop}}

An incoherent superposition of GWs is often decomposed in plane, linearly-polarized, propagating waves as follows~\cite{Maggiore,2013PhRvD..88l4032T, 2010PhRvD..82b2002A, 2017LRR....20....2R}:
\begin{align}
    h_{ij}(t,x) &=\!\!\! \sum_{A=+,\times}\!\int_{-\infty}^{\infty}\!\!\!\!{\rm d}f \!\!\int {\rm d}^2\hat{n} \, \widetilde{h}_A(f,\hat{n})e_{ij}^A(\hat{n})e^{i2\pi f t_{\rm ret}},
\label{eq:gwb_superposition}\\
t_{\rm ret} &= (t-\hat{n}\cdot x/c).
\end{align}
In Eq.~\eqref{eq:gwb_superposition},
 $e_{ij}^A(\hat{n})$ denotes a basis of GW polarization tensors, with indices $A=+,\times$ and
\begin{align}
     e^+_{ij} ({\hat{n}}) &= \hat{u}_i \hat{u}_j - \hat{v}_i \hat{v}_j ,\\
     e^\times_{ij}({\hat{n}}) & = \hat{u}_i \hat{v}_j + \hat{v}_i \hat{u}_j,
\end{align} 
where $\hat{u}$, $\hat{v}$ are unit vectors orthogonal to each other and to the propagation direction $\hat{n}$.
Indices $i,j$ run only over spatial dimensions, as the decomposition is performed in the transverse-traceless gauge.

A stochastic GWB can be characterized by a collection of random variables with an associated probability distribution. 
This is conveniently done in frequency domain by a set of complex amplitudes, corresponding to a plane-wave decomposition of the random field in spacetime. 
Such stochastic processes arise in a number of astrophysical and cosmological contexts~\cite{2022Galax..10...34R}, and are often assumed to be:
\begin{description}
    \item[Ergodic] ensemble averages are asymptotically equivalent to time averages. Henceforth we will refer to them interchangeably, and denote both with  $\langle \cdot \rangle$.
    \item[Stationary] 
    expectation values, e.g.\ mean and covariances of signals, are finite and time-independent. As a consequence, two-point correlations $\langle{h}_A(t){h}_{A'}(t') \rangle$ depend on $t-t'$, only.
    In literature, such signals are frequently referred to as second-order weakly stationary~\cite{2020arXiv200910316M, 1992PhRvD..46.5236F, wiener1930generalized, khintchine1934korrelationstheorie, whittle:1957}. In the Fourier domain, this is equivalently expressed as 
    \begin{equation}
    \label{eq:stationarity}  
    \langle \, \widetilde{h}_A^*(f,\hat{n})\widetilde{h}_{A^\prime}(f^\prime,\hat{n}^\prime)\, \rangle \propto \delta(f-f^\prime) C_{AA^\prime}(\hat{n},{\hat n}^\prime),
    \end{equation}
    where $\delta(f-f^\prime)$ denotes a Dirac-delta in frequency, and $C_{AA^\prime}(\hat{n}, \hat{n}^\prime)$ describes a generic correlation structure across polarizations and propagation directions.

    The proportionality factor in Eq.~\eqref{eq:stationarity} is often referred to as signal spectrum $S_h(f)$, and we choose conventionally to define it over positive frequencies only, thus completing Eq.~\eqref{eq:stationarity} with
    \begin{equation}
        \label{eq:spectrum}  
    \langle \, \widetilde{h}_A^*(f,\hat{n})\widetilde{h}_{A^\prime}(f,\hat{n}^\prime)\rangle = \frac{S_h(f)}{2}C_{AA^\prime}(\hat{n},{\hat n}^\prime).
    \end{equation}
    \item[Gaussian] 
    Higher-order correlation functions can be expressed as a (suitably symmetrized) sum of products of two-point correlations $\langle{h}_A(t){h}_{A'}(t') \rangle$~\cite{Isserlis}.
    In most astrophysical contexts, Gaussianity is a direct consequence of the Central Limit theorem: the sum of many independent and identically distributed random variables asymptotically converges to a Gaussian random variable;
    effectively, this reduces a complete description of the stochastic process to Eq.~\eqref{eq:spectrum}. 
    \item[Isotropic and sky-uncorrelated] 
    No statistical correlations are expected across different wave-propagation vectors, $\hat{n},\hat{n}^\prime$, hence
    \begin{equation}\label{eq:isotropic}
        C_{AA^\prime}(\hat{n},{\hat n}^\prime) = \frac{1}{4\pi}\delta^2(\hat{n},\hat{n}^\prime)C_{AA^\prime}.
    \end{equation}    
    where $\delta^2(\hat{n},\hat{n}^\prime)$ denotes a Dirac-delta on the 2-dimensional unit sphere. 
    \item[Unpolarized] 
    Similarly no correlation is expected between GW polarizations, i.e.
    \begin{equation}\label{eq:unpolarized}
        C_{AA^\prime}(\hat{n},{\hat n}^\prime) = \frac{1}{4\pi}\delta^2(\hat{n},\hat{n}^\prime)\delta_{AA^\prime}.
    \end{equation}
    where $\delta_{AA^\prime}$ denotes a Kronecker delta over polarizations indices.
\end{description}

Under such assumptions, a stochastic background is uniquely characterized by a single function $S_h(f)$. 

If one wishes to describe a certain stochastic signal recorded by a detector, the instrument response must be taken into account, accordingly:
in this work we focus on LISA observation of EMRI GWBs.\@
Given the extragalactic nature of the individual emitters, we assume the background to be unpolarized and isotropic, in the absence of compelling evidence against it (see however Sec.~\ref{subsec:statandgauss} for implications and limitations of the latter assumption).
The observed signal will be affected by typical emission timescales of individual sources, their magnitude with respect to the observation time, the detector response and its sensitivity. 
We elaborate on the above elements in Sec.~\ref{sec:gwb}, following the construction of a few representative populations in Sec.~\ref{sec:emri_pop}.

\section{\label{sec:emri_pop} EMRI population}

We build a selection of EMRI population signals following closely Refs.~\cite{2020PhRvD.102j3023B, 2023PhRvD.108j3039P}.
We first construct a family of populations leading to three representative models out of the twelve presented in Ref.~\cite{2017PhRvD..95j3012B}. Our choice is driven by the findings of Ref~\cite{2023PhRvD.108j3039P}, showing that the EMRI GWB amplitudes are bracketed by \model{8} and \model{12} as extreme scenarios, with \model{1} providing a fiducial, intermediate case.
For each model we generate 10 synthetic catalogs by Monte Carlo sampling from the cosmic EMRI distribution. 
We do so via Monte Carlo sampling of the EMRI population plunging within one year, thus obtaining realizations of EMRI populations across the Universe over 10-year for each of the three models.
We denote them \model{j} with \textsc{j} $= 1, 8, 12$, matching the naming convention used in previous work.
An extensive description of such models is provided therein~\cite{2017PhRvD..95j3012B}. 
We summarize the main features of each model in Table~\ref{tab:models}:
they are characterized by the mass of the MBH, its spin, the effect of cusp-erosion, the $M-\sigma$ relation,  the number of plunges per EMRI $N_p$, and the mass of the CO.
For each one, we construct a catalog of EMRI systems in a simulated Universe.
In doing so, we consider a flat $\Lambda$\textsc{CDM} cosmology ($H_0 = 70 {\rm Km/s/Mpc}$, $\Omega_{{\rm m},0} = 0.3)$. The intrinsic EMRI rate predicted by those models, which we list in Table~\ref{tab:models}, spans over three orders of magnitude, primarily due to the uncertain number of plunges and the limited constraints on the MBH mass distribution at its low end. A list of primary MBH masses $M$ and spins $a$, along with a redshift $z$ for each event is given.

In Ref.~\cite{2017PhRvD..95j3012B}, the mass $\mu$ of the CO was fixed to $10$ or $30 {\rm M}_{\odot}$.
Here, we instead sample it from a distribution inspired by the population of compact binary mergers observed by GW ground-based detectors. 
In reality, it is likely that COs in EMRIs follow a more top-heavy mass distribution compared to that of coalescing compact binaries. 
This is because EMRIs form in galactic nuclei, where the more massive COs are expected to cluster toward the center due to dynamical processes such as relaxation and mass segregation \cite{2006ApJ...649...91F,2006ApJ...645L.133H}.
However, in the absence of robust observational constraints on the CO mass distribution in galactic nuclei, we just opt here to introduce a reasonable scatter in the underlying mass distribution to explore its role. 
A more realistic population choice goes beyond the scope of this work and is left for future investigation.

\begin{figure}[t]
    \centering
    \includegraphics[scale=0.5]{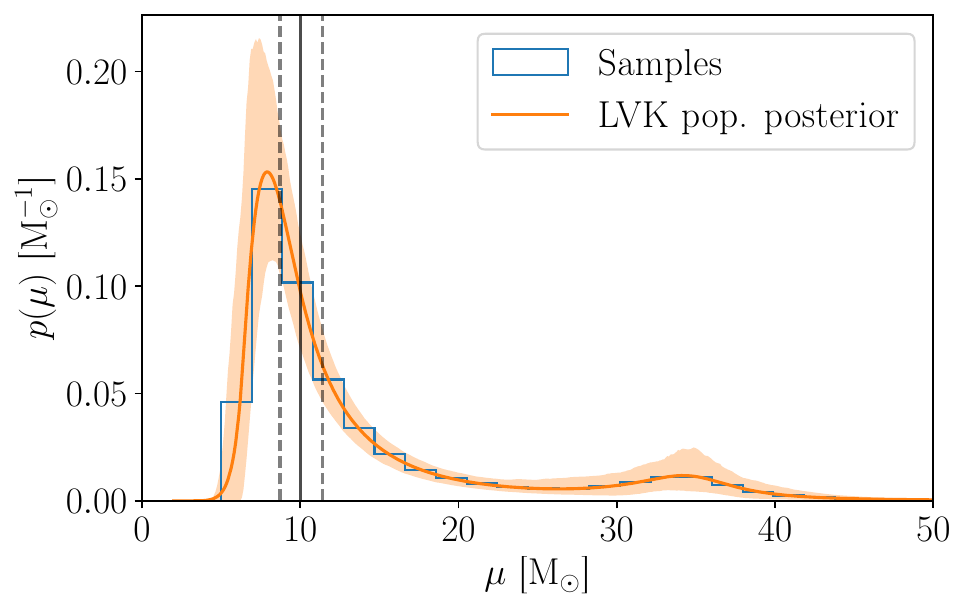}
    \caption[\ac{PDF} of \ac{SOBH}.]
    {Distribution of stellar-mass black hole based on current gravitational wave detections~\cite{2023PhRvX..13a1048A}. The orange-shaded area represents the overall posterior distribution, while the solid orange line shows our selected reference model. Solid and dashed black lines denote the posterior on the median, $\mu_{\rm med} = 10^{+1.4}_{-1.3} M_{\odot}$, which is in large agreement with the fixed $\mu = 10 M_{\odot}$ model employed in Ref.~\cite{2023PhRvX..13a1048A}. In blue are highlighted the samples from the chosen distribution used in the analysis.\label{fig:m2_pdf}}
\end{figure}

In the third Gravitational Wave Transient Catalog (\textsc{GWTC-3})  observations of the Advanced LIGO~\cite{2015CQGra..32g4001L} and Virgo~\cite{2015CQGra..32b4001A} detectors from the first three observing runs (O1, O2, and O3, respectively) are collected~\cite{2023PhRvX..13d1039A}. 
Among those, 69 confident BBH events are identified based on their significativity to perform population inference~\cite{2023PhRvX..13a1048A} with a variety of mass-distribution models.
As a reference for our CO mass distribution, we choose the point-wise median of the marginal population posterior on $\mu$ \textsc{PowerLaw+Peak} model.
We highlight here that our extension remains in excellent agreement with the former choice of fixed-CO mass in Ref.~\cite{2017PhRvD..95j3012B}, as shown in Fig.~\ref{fig:m2_pdf}. 
The median of our chosen CO mass distribution (solid line) is very close to the nominal value of $10 M_{\odot}$, and is in good agreement with the variability inferred in Ref.~\cite{2023PhRvX..13a1048A} (dashed lines). 
For reference, we also show the overall population posterior (orange shaded regions) and our chosen reference model (orange solid line).

\begin{table*}
    \footnotesize
    \centering
    \renewcommand{\arraystretch}{1.2} 
    \begin{tabular}{c|lccc|lcc}
    \hline\hline
    Model&  MBH Mass & MBH spin & Cusp Erosion & $M-\sigma$  & $N_p$ & Rate [${\rm yr}^{-1}$]\\
    \hline
    \model{1}& Barausse12& a98& yes& Gultekin09& 10&  1600\\
    \model{8}& Barausse12& a98& yes& Gultekin09& 100&  180\\
    \model{12}& Barausse12& a98& yes& Gultekin09& 0&  20000\\
    \hline\hline
    \end{tabular}
    \caption{EMRI population models considered in this study. 
    Column 1 denotes the label of each model. 
    The three selected models, a subset of those presented in Ref.~\cite{2017PhRvD..95j3012B} share the MBH mass distribution labeled as Barausse12~\cite{2012MNRAS.423.2533B, 2014ApJ...794..104S, 2015ApJ...806L...8A, 2015ApJ...812...72A} (column 2), the MBH spin model labeled as a98 denoting distribution peaked to $a = 0.98$ (column 3), the inclusion of the cusp erosion effect following MBH binary mergers (column 4), and the $M-\sigma$ relation labeled as Gultekin09~\cite{2009ApJ...698..198G} (column 5). By contrast, they have different plunges-to-EMRIs ratio (column 6). For reference, we show in column 7 the total number of EMRIs occurring in a year up to $z = 4.5$.}\label{tab:models} 
\end{table*}

To fully characterize the EMRI signal we follow the convention adopted in~\cite{2021PhRvD.104f4047K, 2017PhRvD..96d4005C}. Fourteen parameters are required: nine intrinsic ones describing the phase evolution which, together with the five extrinsic ones define the GW amplitude.
In addition to the previously introduced MBH mass $M$, CO mass $\mu$, dimensionless primary Kerr spin parameter $a$, the intrinsic ones are completed by the initial semi-latus rectum $p_{0}$, eccentricity $e_{0}$, inclination angle $\iota_{0}$, and the initial azimuthal, polar, and radial phases $\Phi_{\phi,0}, \Phi_{\theta,0}, \Phi_{r,0}$. 
The extrinsic parameters are instead the luminosity distance $d_L$, the sky position in polar- and azimuthal-angle in the solar system barycenter frame $\theta_S$ and $\phi_S$, and the orientation of the MBH spin polar- and azimuthal-angle with respect to the ecliptic $\theta_K $ and $ \phi_K$. 
In summary, each source is fully described by the set of 14 parameters\footnote{We neglect here the (small) effect on the signal induced by the spin of the orbiting CO, which would require three additional parameters for its magnitude and orientation.} $\boldsymbol{\theta} = \{M, \mu, a, p, e, \iota, d_L, \theta_S, \phi_S, \theta_K, \phi_K, \Phi_{\phi,0}, \Phi_{\theta,0}, \Phi_{r,0}\}$.
We provide below details on their construction, closely following the approach in Refs.~\cite{2023PhRvD.108j3039P,2020PhRvD.102j3023B}.

\begin{itemize}
\item Without entering into the details of EMRI formation channels, we assume that EMRIs form via relaxation-driven capture of COs in spherical nuclear clusters. 
Therefore, we assign a cosine of the inclination angle, $\cos \iota$, by randomly drawing values from a uniform distribution between $[-1,1]$.
This is a convenient parameterization, in that \(\iota\) spans the entire interval \([0,\pi]\), with prograde (retrograde) orbits corresponding to \(0 \leq \iota \leq \pi/2\) (\(\pi/2 \leq \iota \leq \pi\)). 
\item We assume isotropic distributions for sky positions \((\theta_S,\phi_S)\) and spin orientations \((\theta_K,\phi_K)\). 
\item Similarly, we assume uniform distributions in $[0, 2\pi]$ for the three initial phases $\Phi_{\phi,0}, \Phi_{\theta,0}, \Phi_{r,0}$.
\item We use our knowledge of the MBH mass \(M\) and spin \(a\) to calculate the radius at the Kerr innermost stable circular orbit $r_{\rm ISCO}$. This defines the starting point of the EMRI evolution, described below.
\item Following Ref.~\cite{2017PhRvD..95j3012B}, for each system, we draw the eccentricity at the last stable orbit \(e_p\) from a flat distribution in the range \([0; 0.2]\).
\item We integrate the system orbital elements backward in time from $r_{\rm ISCO}$ for \(T_{\rm back}\). The latter is obtained using the linear scaling law \(T_{\rm back} = 20(\frac{M}{10^4 \rm M_{\odot}})\rm yr\). EMRIs featuring low-mass primaries predominantly emit GWs within the LISA band during the last few years of their evolution. In contrast, those involving higher MBH masses emit several hundreds of years before plunge.
We randomly sample $ N_{\rm back} = \floor*{(T_{\rm back}/10 {\rm yr})}$ points within the interval $[0;T_{\rm back}]$ to represent different EMRI evolutionary stages. 
The renormalization factor of $10 {\rm yr}$ compensates for the Montecarlo catalog construction described earlier, to avoid overestimating the EMRI rate.
For each of the $N_{\rm back}$ sampled points, we compute the semi-latus rectum $ p_0 = a_{\rm sm,0}(1-e_0^2)$ using the initial semi-major axis $a_{\rm sm,0}$ and eccentricity $e_0$.
\end{itemize}
Following the approach above, we create \(N_{\rm back}\) synthetic copies of each EMRI from the catalogs listed in Table~\ref{tab:models}, sharing the same inclination angle, redshift, mass, and spin. 
Despite this limitation, the 10-year EMRI catalogs contain thousands of events covering a broad range of MBH masses and redshifts. 
As described in~\cite{2020PhRvD.102j3023B}, negligible biases are introduced in the foreground computation through such a procedure.

\section{\label{sec:data} LISA Data processing}

\subsection{\label{sec:gwb} Signal construction}

The binary systems we are modeling are typically characterized by mass ratios $\epsilon = \mu/M < 1$ between $\epsilon = 10^{-3}$ and $10^{-6}$.
Therefore, it is possible to accurately derive their GW waveform, treating the small CO as a perturbation to the background Kerr metric associated to the primary MBH~\cite{2019RPPh...82a6904B, 2012PhRvL.109e1101P, 2022hgwa.bookE..38P, 2009CQGra..26u3001B}. 
Small mass-ratio waveforms built directly within the gravitational self-force (GSF) formalism are very accurate, though computationally expensive. 
This renders them impractical for data analysis purposes, which instead require waveform evaluation times below a few hundred milliseconds (see~\cite{2023PhRvL.130x1402W, 2024PhRvD.109l4048B} and references within). 

The goal of \texttt{FastEMRIWaveforms} package~\cite{2021PhRvD.104f4047K} is to generate GSF-based waveform models that can be evaluated in $\lesssim 10$ milliseconds. 
Currently, only eccentric Schwarzschild models with leading-order terms in the mass-ratio (adiabatic) GSF information are publicly available. 
Since all astrophysical MBHs at the center of galactic nuclei must rotate due to the conservation of angular momentum, and the interaction with the environment, the eccentric Schwarzschild model is unsuitable for building realistic astrophysical backgrounds. 
For our purposes, we require fully generic EMRI waveforms that encapsulate the rotation of the primary black hole and both eccentric and inclined orbits. 
No such rapid-to-evaluate GSF-based EMRI waveforms exist for this class of orbits. 

Instead, we rely on a specific family of approximate generic-orbit EMRI waveforms, referred to as ``kludges''. 
The kludge approach is to build approximate EMRI waveforms that cover the entire parameter space while capturing the realistic behavior of generic orbits under radiation reaction effects. 
There are three main types of kludges available in the literature. 
The Analytic Kludge (AK)~\cite{2004PhRvD..69h2005B}, Numerical Kludge (NK)~\cite{2007PhRvD..75b4005B}, and the 5PN Augmented Analytic Kludge (5PN-AAK)~\cite{2021PhRvD.104f4047K}. 
The 5PN-AAK model evolves the orbital parameters using a fifth order post-Newtonian expansions for small eccentricities. 
The (quadrupolar $l = m = 2$) amplitudes are constructed using a weak-field Peters and Matthews approximation~\cite{1963PhRv..131..435P} to the metric perturbation. 
In our work, we will use the 5PN-AAK waveform model integrated into the \texttt{FastEMRIWaveforms} package, following Refs.~\cite{2021PhRvL.126e1102C, 2004PhRvD..69h2005B}.

After an EMRI signal is constructed, we need to compute the response of the LISA instrument to the incoming gravitational wave. Our analysis employs the most recent LISA noise PSD from the SciRDv1 model~\cite{LISAsr:18aa}.
We adopt the approximation of 1st generation ``noise-orthogonal'' time-delay-interferometric (TDI) variables $A$, $E$, and $T$, suitable for equal and static LISA armlengths~\cite{1999ApJ...527..814A, 2021LRR....24....1T, 2022arXiv220201124H}.
This approach implements the TDI technique, crucial to suppress the dominant laser phase noise in LISA~\cite{2021LRR....24....1T}. 
For this purpose, we use the \textsc{Python} package \texttt{FastLISAResponse} available in Ref.~\cite{2022PhRvD.106j3001K}.
Our analysis primarily focuses on the $A$ and $E$ channels, as the $T$ channel is significantly less sensitive to GWs by construction~\cite{1999ApJ...527..814A}.
Having introduced TDI variables, the computation of the single source signal-to-noise-ratio (SNR) reads 
\begin{equation}
    \rho = \sqrt{\sum _k (h_k|h_k)_k},
    \label{eq:snr_tdi}
\end{equation}
where $k$ refers to the optimal TDI variables $k=(A, E)$, and the inner product between two-time series $a(t)$ and $b(t)$ is expressed as 
\begin{equation}
    (a|b)_k = 4{\rm Re}  \int_{0}^{\infty} {\rm d}f  \, \frac{\widetilde{a}^*(f) \widetilde{b}(f)}{S_{n,k}(f)},
\label{eq:inner_prod}
\end{equation}
where $S_{n,k}(f)$ denotes the power-spectral density (PSD) of each TDI variable $k$.

\subsection{\label{sec:gwb_eval} Building EMRI backgrounds}
After constructing the EMRI parameters as shown in Sec.~\ref{sec:emri_pop} we compute the plus- (cross-) polarized strain $h_+(t)$ ($h_\times(t)$) for each source using the 5PN-AAK waveform, and evaluate the individual TDI signals $A(t)$, $E(t)$, and $T(t)$. 
This process has variable computational costs, strongly dependent on the specific source parameters (predominantly its initial eccentricity and semilatus-rectum).
To minimize the computational time required for the signal generation, we follow two main strategies:
(i) all calculations are performed on NVIDIA A100 GPUs, as CPU evaluation would be unfeasible. 
We use the \texttt{FastEMRIWaveforms} package for GPU-accelerated EMRI waveform generation and GPU-repurposed libraries from \texttt{FastLISAResponse} for TDI variable calculation~\cite{2022PhRvD.106j3001K}. This GPU implementation reduces initial runtimes by one to two orders of magnitude;
(ii) following Ref.~\cite{2020PhRvD.102j3023B}, we remove the faintest sources from each catalog.
We compute an approximate SNR ($\rho_{\rm AK}$) for each source using an inclination-polarization averaged version of the AK waveform. This method, computationally inexpensive (below a millisecond per source evaluation), uses a 2PN approximation for the orbital parameters and the Peters-Mathews approximation for the GW amplitude. We select sources with $\rho_{\rm AK} > 1$ for the subsequent background generation. 
As shown in Table~\ref{tab:n_source}, this reduces the number of sources by $\sim 97\%$. However, Ref.\cite{2023PhRvD.108j3039P} demonstrates that this approach decreases the background SNR by only $\sim 5\%$. In Sec.~\ref{sec:discussion} we discuss the accuracy of this approximation, and its impact on the EMRI foreground SNR. 

\begin{table}[t]
\footnotesize
\centering
\begin{tabular}{p{1cm}|p{1.5cm}p{1.5cm}}
\hline\hline
Model& $N_{\rm start}$& $N_{\rm final}$ \\ 
\hline
\model{1} & 1217952  & 26932  \\ 
\model{8} & 124968   & 3209   \\ 
\model{12} & 21315202& 319309 \\ 
\hline\hline
\end{tabular}
\caption{Selection of sources across the three catalogs considered. The initial (final) number of sources retained in each EMRI catalog is listed in the second (third) column, the threshold for retention being $\rho_{{\rm AK},{\rm th}} = 1$.\label{tab:n_source}}
\end{table}

To define the dataset, we make several key modeling choices. 
We consider a LISA mission duration of $T_{\rm OBS} = 4$ years, resulting in a frequency resolution of $\Delta f = 1/T_{\rm OBS} \approx 8 \times 10^{-9} \text{ Hz}$. The cadence is set to $\Delta t = 20 \text{ s}$, corresponding to a maximum non-aliased frequency of $f_{\text{max}} = 1/(2\Delta t) = 0.025 \text{ Hz}$. While a realistic data stream will likely have a higher sampling rate of approximately 0.25 s, we note that for $f > 0.01 \text{ Hz}$, the LISA noise budget begins to increase, and the EMRI foreground brightness starts to decrease \cite{2023PhRvD.108j3039P}. 
Thus, our choice of $f_{\rm max} = 0.025{\rm Hz}$ allows us to characterize the statistical properties within a region of primary interest for LISA observations. 

The EMRI background signals are then obtained as follows:
\begin{enumerate}
    \item  We re-compute the two strain amplitudes associated with each source in a catalog 
    \begin{equation}
    {\left\{ h^{(i)}_{+}(t), h^{(i)}_\times(t)\right\}}_{i=1}^{N_{\rm final}}
    \end{equation}
    using 5PN-AAK waveform, where $N_{\rm final}$ is the total number of sources in the catalog, after the initial selection, as listed in Table~\ref{tab:n_source}.
    \item We then generate the noise-orthogonal TDI variables $A$, $E$ and $T$, using the \texttt{FastLISAResponse} package
    \begin{equation}
    {\left\{A^{(i)}(t), E^{(i)}(t), T^{(i)}(t)\right\}}_{i=1}^{N_{\rm final}}.
\end{equation}
    \item 
    We evaluate individual source accurate SNR $\rho$, as in Eq.~\eqref{eq:snr_tdi},  and separate them into two classes: ``resolvable'' and ``unresolvable'', characterized by $\rho \ge 20$ and $\rho < 20$, respectively. Our choice is consistent with values assumed in literature to establish an EMRI detection ~\cite{2017PhRvD..95j3012B, 2020PhRvD.102j3023B, 2023PhRvD.108j3039P, 2024PhRvD.109l4048B}. 
    \item For each source specific TDI data stream, we sum the individual TDI channel contribution to build a single data stream that is a collection of background sources.
\end{enumerate}

Following this procedure, we construct the EMRI backgrounds for models \model{1}, \model{8}, and \model{12} in the time domain thus obtaining three global time series for each catalog 
\begin{align}
    A(t)&=\sum_{i\in \mathcal{B}} A^{(i)}(t), \label{eq:Achannel}\\
    E(t) &=\sum_{i\in \mathcal{B}} E^{(i)}(t), \label{eq:Echannel}\\
    T(t) &=\sum_{i\in \mathcal{B}} T^{(i)}(t)\\
    \mathcal{B} & = \left\{j\mid {\rho}_j < 20, j = 1,\dots,N_{\rm final}\right\} \label{eq:Tchannel}
\end{align}
In computing the single source SNR $\rho_j$, we consider two components for the noise spectral density $S_n(f)$: 
the LISA instrumental noise $S_{\text{instr}}(f)$ and the confusion noise arising from the superposition of unresolved Galactic binaries $S_{\text{gb}}(f)$, such that $S_n(f) = S_{\text{instr}}(f) + S_{\text{gb}}(f)$. 
A more realistic approach would be to perform an Iterative Foreground Estimation (IFE) method, as e.g. described in Ref.~\cite{2021PhRvD.104d3019K}. 
This entails iteratively processing a catalog and update the reference noise $S_n(f)$ to include unresolved sources from the catalogue itself, to then use it to estimate each source $\rho$. Though an approximate algorithm with respect to existing global fit pipelines~\cite{2024arXiv240504690K,2023PhRvD.107f3004L,2024PhRvD.110b4005S}, it has been shown to be a sufficiently accurate proxy for them. 
We defer a discussion on the impact of our choice after the examination of the results, in Section~\ref{sec:discussion}.

Since the Fourier transform is a linear operator, we readily obtain the signals in frequency domain $\tilde{A}(f),\tilde{E}(f)$. 
The associated spectra are shown in Fig.~\ref{fig:gwb_A} for channel $A$.
To characterize each GWB brightness, we will use its total SNR $\rho_{{\rm gwb}}$, which is obtained as the sum in quadrature of the SNRs $\rho_{{\rm gwb},i}$ in the $A$ and $E$ channels, 
following Refs.~\cite{2017LRR....20....2R, 2019JCAP...11..017C}, through:

\begin{equation}
{\rho}_{{\rm gwb},i} = \sqrt{T_{\rm OBS} \int^{\infty}_{0} {\rm d}f{\left({\frac{S_{{\rm gwb},i}}{S_{{\rm n},i}}}\right)}^2},
\label{eq:snr_gwb}
\end{equation}
where $i$ label $A$ or $E$, and $S_{{\rm gwb},i}$ denotes the corresponding channel PSD.

\section{\label{sec:Ray_test} Rayleigh test}

We now perform the statistical test on the datastreams constructed following Eq.~\eqref{eq:Achannel} and Eq.~\eqref{eq:Echannel}. 
We take inspiration from the literature available for ground-based detectors~\cite{Rayleigh,2022arXiv221015634A}, where such a test is employed for data-quality diagnostics, e.g. to detect the presence of glitches.
The null-hypothesis we test against is that of an ergodic, zero-mean stationary Gaussian process. 
We further assume the signal to be second-order stationary.
Hence, in time-domain the timeseries represent a fair draw from a Gaussian distribution
\begin{equation}
    x(t) \sim {\cal N}(0, \Sigma(t,t^\prime)).
\end{equation}

Representing the Fourier transform of a time-domain data $x(t)$ as, 
\begin{equation}
    \tilde x(f) = \int^\infty_{-\infty}{\rm d}t\ x(t) e^{-{\rm i}2\pi ft} ,
\end{equation}
and being the Fourier transform a linear operator, the process is readily cast in frequency domain as:
\begin{equation}
    \Re x(f), \Im x(f) \sim {\cal N}(0, S(f))\label{eq:realpart}
    \end{equation}

where $S(f)$ denotes the spectrum of the signal and $\Re,\Im$ denote the real and imaginary part of a complex number, respectively.
For a stationary stochastic process, the PSD is defined as 
\begin{equation}
S(f)=\lim_{T\rightarrow +\infty}\frac{1}{2T} \left\langle \left| \int_{-T}^{+T}{\rm d}t\ x(t) e^{- {\rm i} 2\pi f t}\right|^2\right\rangle .
    \label{eq:stoch_proc_spectrum}
\end{equation}

The absolute value of the complex variable $x(f)$ is therefore distributed as follows
\begin{equation}
    \left| x(f) \right| := \sqrt{\Re x(f)^2 + \Im x(f)^2} \sim {\rm Rayleigh}(\sigma(f))\label{eq:rayleighdistro}
\end{equation}
Similarly, the squared norm of the Fourier variable is distributed as follows
\begin{equation}
    \left| x(f) \right|^2 \sim \Gamma(1,2\sigma(f))\label{eq:gammadistro}
\end{equation}
where $\Gamma(k,\theta)$ denotes the Gamma distribution and $k,\theta$ are the positive shape and scale parameters, respectively.
The Gamma distribution has mean $k\theta$ and variance $k\theta^2$, respectively.
The test is constructed following Eq.~\eqref{eq:gammadistro} by taking the ratio between the standard deviation and mean of the Gamma distribution.
Under the null-hypothesis, the variable is identically one across all frequencies
\begin{equation}
    \rho_{\left[x\right]}(f) = 
    {\frac
    {\sqrt
    {\left\langle \left(\left|x(f)\right|^2 - \left\langle \left|x(f)\right|^2\right\rangle\right)^2
    \right\rangle}
    }
    {\left\langle {\left|x(f)\right|}^2\right\rangle
    }
    }
    =\frac{2\sigma(f)}{2\sigma(f)} = 1
    \label{eq:nullhyp}    
\end{equation}
where $\rho$ denotes the operator acting on the process $x$.
We will denote the expected value one as, $\rho_{[x],\rm exp}$.

The Rayleigh test is then a statistical test of the stationarity and Gaussianity hypotheses that the \emph{coefficient of variation} (standard deviation over mean) of the signal FFT has the value expected for the corresponding Rayleigh random variable.
An equivalent statistic is that implemented by LIGO and Virgo where the squared norm, distributed like a Gamma variable~\cite{DiRenzo}, is considered. The test is practically carried out by constructing estimators for the random quantities in Eq.~\eqref{eq:nullhyp}.
To obtain multiple samples we leverage the process ergodicity and suitably chunk the data. The denominator is evaluated through Welch's PSD estimator~\cite{1967ITAE...15...70W} while the numerator is obtained through FFT, and represents a measure of the signal's statistical properties variability.

Critical values for the test statistic can be obtained under the null-hypothesis, i.e. for a perfectly Gaussian, stationary signal of the same (finite) duration of our GWB datastreams.
Asymptotically, for a finite number $N$ of samples (which in this context are to be regarded as the chunks), the test is distributed as a ${\cal N}\big(1, 1/2\sqrt N\big)$. 
Ref.~\cite{2024arXivDWD} presented a simplified example demonstrating how violations of stationarity and Gaussianity are detected using such test. 
In Sec.~\ref{sec:discussion} we use it to characterize the Gaussianity and stationarity of EMRI GWBs.

\section{\label{sec:discussion} results}

\subsection{EMRI background spectra and SNR \label{subsec:emrispectra}}

\begin{figure}[t!]
    \centering
    \includegraphics[width=\columnwidth]{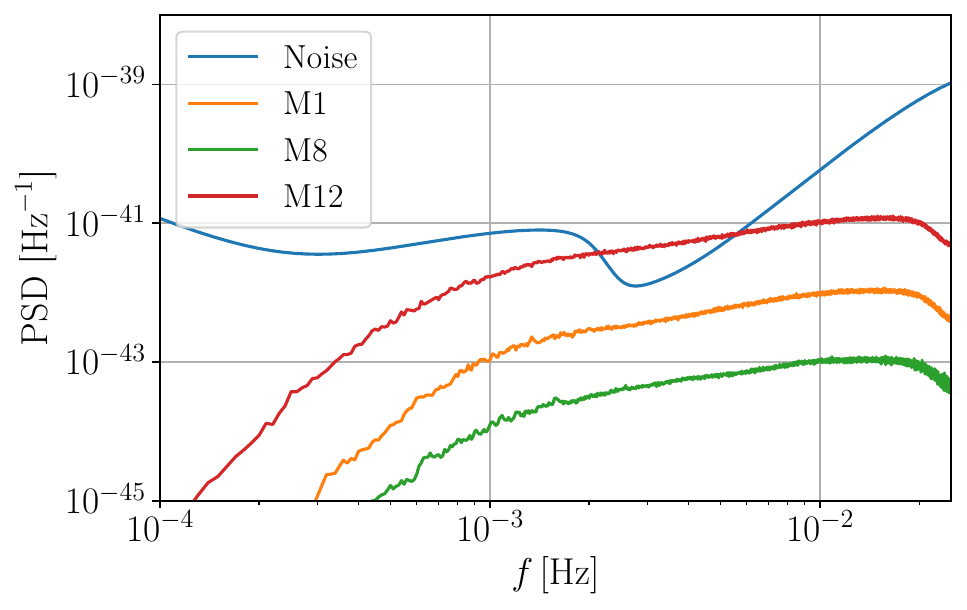}
    \caption{GWB power spectral densities for the three reference EMRI models, as observed in the LISA $A$ channel. The solid blue line represents the LISA instrumental noise, together with the Galactic confusion noise. The red (orange, green) line corresponds to model \model{12} (\model{1}, \model{8}), as presented in Sec.~\ref{sec:emri_pop}, and following the construction outlined in~\ref{sec:gwb_eval}. 
    Corresponding SNRs and number of resolvable sources are listed in Table~\ref{tab:dettable}.\label{fig:gwb_A}}
\end{figure}

Using the populations constructed in Sec.~\ref{sec:emri_pop}, we bracket our predictions using models \model{12}, \model{8}, and \model{1}.
We show the resulting GWB spectra in Fig.~\ref{fig:gwb_A}.
The predicted EMRI background spans three orders of magnitude in power spectral density amplitude across all frequencies relevant for LISA, consistently with the rates reported in Ref.~\cite{2017PhRvD..95j3012B} and the spectra obtained in Ref.~\cite{2023PhRvD.108j3039P}.
Model \model{12} is found to be louder than the LISA noise level. 
On the other hand, \model{8} is significantly below the sensitivity curve, while \model{1} lies in between.
Nonetheless, the dependence on $T_{\rm OBS}$ in Eq.~\eqref{eq:snr_gwb} yields large signal-to-noise ratios for all models considered.
Table~\ref{tab:dettable} summarizes them, alongside the number of detections for each model considered.
Following results in literature, we adopt a GWB detection threshold of $\rho>10$~\cite{2013PhRvD..88l4032T,2024arXiv241210468P}, hence classify all three models considered as largely detectable over the target mission duration of $T_{\rm OBS}=4 {\rm yr}$.

Despite the large agreement with Ref.~\cite{2023PhRvD.108j3039P}, we highlight here a few differences in implementation potentially affecting the results above.
In this study, we allowed the mass of the CO to vary, rather than fixing it at $10 {\rm M}_{\odot}$.
In addition, we consider the galactic confusion noise in the LISA noise budget appearing in Eq.~\eqref{eq:snr_tdi} and Eq.~\eqref{eq:snr_gwb}.
This results in fewer individual detections and systematically higher GWB SNRs observed in models \model{1} and \model{8}.
More importantly, we do not employ an IFE algorithm, obtaining consequently a smaller number of unresolvable sources and a fainter foreground.
As shown in Ref.~\cite{2023PhRvD.108j3039P}, such an algorithm yields a drastically decreased number of resolvable sources for foregrounds above the LISA noise level. 
This is in fact the case of model~\model{12}.
Conversely, foregrounds below the noise level are less affected by this difference. 

\begin{table}[!ht]
\centering
\begin{tabular}{c|ccc}
\hline\hline
Model & $N_{{\rm final}}$ & Detections & ${\rm \rho}_{{\rm gwb}}$\\
\hline
\model{1} & 26932  & 522  & 311 \\
\model{8}  &  3209  &  64  & 38 \\
\model{12} & 319309 & 5909 & 3684 \\
\hline\hline
\end{tabular}
\caption{Summary of results for each EMRI catalog. The first column indicates the model. The second column reports the number of sources in each EMRI catalog used for each background computation. 
The third one shows the number of resolvable sources ($\rho > 20$). 
The GWB SNR is listed in the fourth column. 
Signals are generated with AAK waveforms described in Sec.~\ref{sec:data} for a nominal 4 year-long LISA observation time.}
\label{tab:dettable}
\end{table}

\subsection{EMRI backgrounds stationarity and Gaussianity}
\label{subsec:statandgauss}

\begin{figure*}[ht!]
    \centering
    \includegraphics[width=\textwidth]{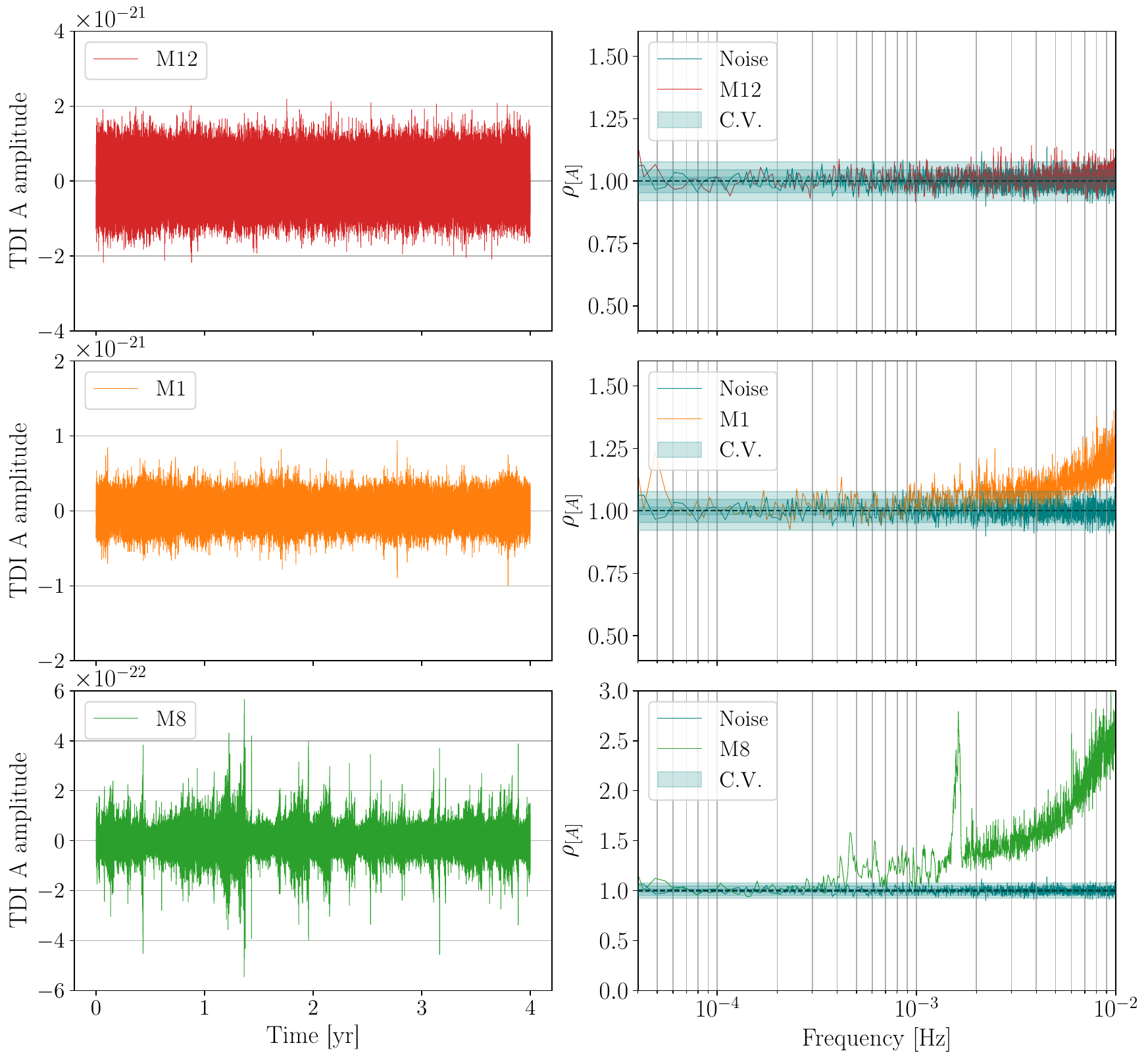}
    \caption{Rayleigh test applied to the constructed EMRI foregrounds. Left column: EMRI GWB in the time domain A channel for models \model{12} (top), \model{1} (middle), and \model{8} (bottom), to notice the different scales on the y-axis. Right column: Rayleigh statistics as a function of frequency in logarithmic scale. The test applied to a stationary and Gaussian stochastic time series is shown by the teal region. Within this region, the expected value $\rho_{[A],\rm exp}=1$ is shown as a black dashed line, while inner (middle, outer) teal-shaded bands denote the critical values for the null-hypothesis rejection, at $1\sigma$ ( $2\sigma$, $3\sigma$) confidence. As mentioned in Sec.~\ref{sec:Ray_test} these are asymptotic values constructed for a finite number of realizations of a stationary Gaussian process.} \label{fig:gwb_ray}
    \end{figure*}

\begin{figure}[ht!]
    \centering
    \includegraphics[width=\columnwidth]{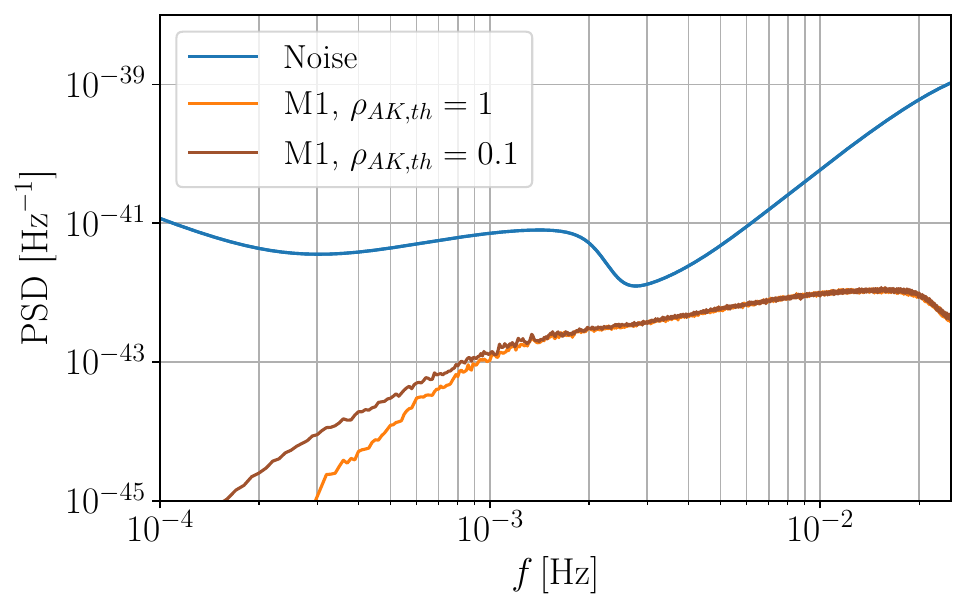}
    \caption{GWB power spectral densities in the LISA A channel for model \model{1} with two different preliminary cuts in the populations, $\rho_{AK,{\rm th}} = 1$ ($\rho_{AK,{\rm th}} = 0.1$) solid orange line (solid brown line). Solid blue line represents the LISA instrumental noise, together with the Galactic confusion noise model. The inclusion of fainter sources in the analysis populates the low-frequency band. 
    \label{fig:gwb_A_1_01}}
    \end{figure}

\begin{figure*}[ht!]
\centering
\includegraphics[width=\textwidth]{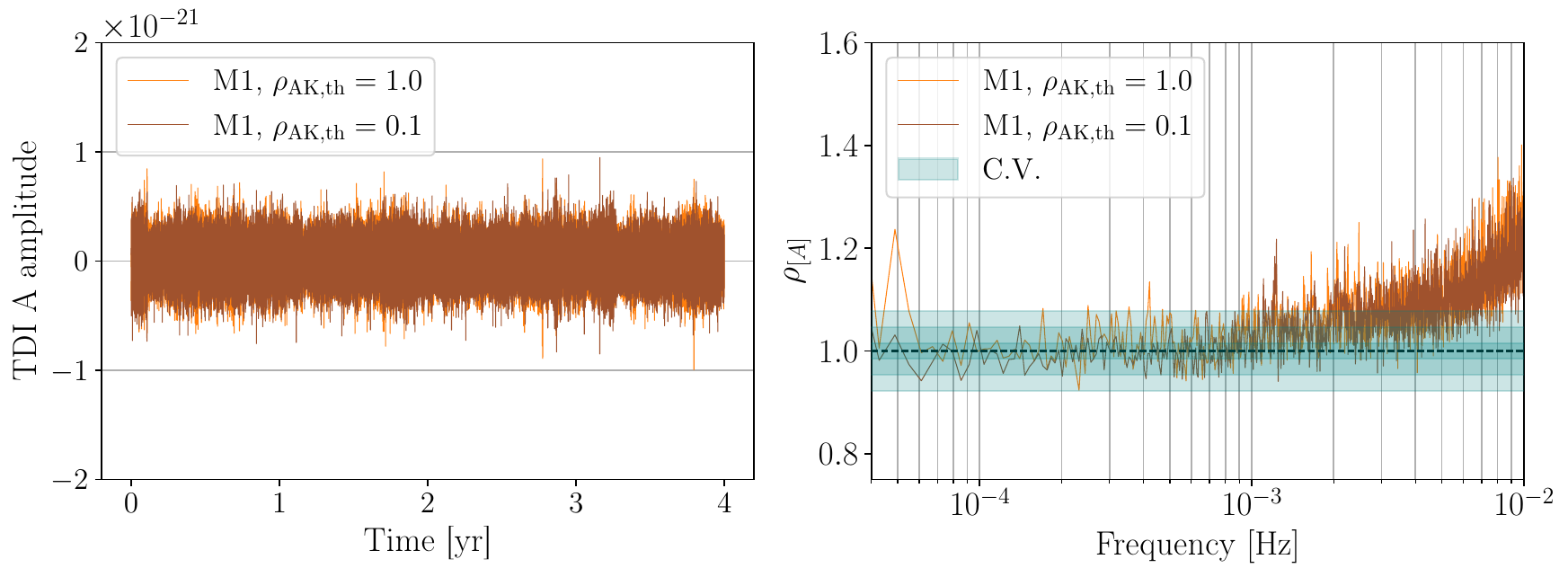}
\caption{Rayleigh test applied to the constructed EMRI foregrounds model \model{1} with varying downselection SNR thresholds $\rho_{\rm AK, tr} = 1 \quad (0.1)$ in orange (brown).
Left panel: EMRI GWB in the time domain A channel; right panel: Rayleigh statistics as a function of frequency in logarithmic scale. In the right panel, the expected value $\rho_{[A],\rm exp}=1$ is shown as a black dashed line, and the shaded blue area represents the Rayleigh test's confidence intervals.
Adding faint sources does not significantly change the test statistics, as the loudest sources dominate the time series.
\label{fig:gwb_ray_M1}}
\end{figure*}

To assess the stationarity and Gaussianity of the EMRI backgrounds, we employ the Rayleigh test introduced in Section~\ref{sec:Ray_test}.
This test serves as a detection statistic for time-series non-Gaussianities and non-stationarities.
Our catalog construction enforces isotropically distributed sources. Therefore, a GW detector receives an overall signal (and couples to it) whose spectral contributions arise equally from each direction.
This assumption is expected to hold for cosmological backgrounds but may be partially or entirely violated for astrophysical ones. 
As described in Eq.~\eqref{eq:gwb_superposition} of Sec.~\ref{sec:forestatprop}, stochastic signals are often modeled as the superposition of infinitely many signals overlapping in time or frequency domain.
However, for this limit to be a suitable approximation, the central limit theorem must hold, justifying the use of a Gaussian likelihood model for inference~\cite{2017LRR....20....2R}: for each given frequency and sky-location intervals, the number of sources contributing individually must be $\gg 1$. 
Failure in satisfying this constrain implies different degrees of deviations from Gaussianity and stationarity.
Often, the resulting signal exhibits a ``popcorn-like'' behaviour~\cite{2024arXiv241017180R}.
While the frequency content is non-trivial to characterize due to the broadband nature of each signal, the sky distribution can be readily tested via catalogs inspection. We do so by performing
Upon cross-checking our hypothesis, we observe through a simple angular multipole decomposition and find that \model{8} (\model{1}) yields GWB power at $\ell>1$ relative to the monopole ($\ell =0$) 100 (10) times larger than \model{12}. This is confirmed by the increasing degree of popcorn-like behaviour for the timeseries we construct below and present in Fig.~\ref{fig:gwb_ray}.

Following the common practice for statistical tests on time series when only a single realization is available, we invoke ergodicity (i.e., the equivalence in distribution between ensemble averages and time averages) as a working hypothesis.
We divide the time series into 3000 chunks, resulting in a frequency resolution of $\Delta f \sim 10^{-5} {\rm Hz}$. 
This value is small enough to consider individual chunks' spectra approximately stationary, provides suitable resolution in the ${\rm mHz}$ band, and yields a large enough set of samples for ensemble average estimators with small variances.
Fig.~\ref{fig:gwb_ray} (right column) shows the results of the Rayleigh test statistic for the three models considered.
We present the test statistics $\rho_{[A]}$ as a function of frequency for the TDI channel A;
depending on the considered EMRI catalogs, approaching higher frequencies ($f \gtrsim 1 {\rm mHz}$), the test shows varying degrees of deviation from the expected value $\rho_{[A],\rm exp}=1$ (shown as a black line in each panel), suggesting the presence of either non-Gaussianities or non-stationarities at such frequencies.
   
Upon closer inspection of each model, we discern the test response to variations in the number of sources in the catalogs:
\begin{itemize}
\item \model{8} (Fig.~\ref{fig:gwb_ray}, bottom row): the time-domain GWB shows a time-modulation due to the low number of sources in the catalog contributing to the foreground ($N_{\rm final} = 3209$). 
This is expected to be a source of deviation from the null hypothesis, i.e. stationarity and Gaussianity. 
The test exhibits small fluctuations at low frequencies and more substantial ones at $f \gtrsim 1 {\rm mHz}$.
See also Ref.~\cite{2024arXivDWD} for a discussion on non-Gaussian signals as non-stationarity mimickers for tests involving ergodicity.
\item \model{1} (Fig.~\ref{fig:gwb_ray}, middle row): the larger number of sources ($N_{\rm final} = 26932$) corresponds to a smaller time-modulation compared to \model{8}. 
The test fluctuations at low frequencies disappear, but deviations at high frequencies exceeding $\sim 1\, {\rm mHz}$ are still noticeable, indicating a stationarity or Gaussianity violation, albeit much smaller than M8.
\item \model{12} (Fig.~\ref{fig:gwb_ray}, top row): having the largest number of sources in the catalog ($N_{\rm final}=319309$), the high-frequency range of the GWB is more densely populated, making the test violation almost negligible at the target highest significance of $3\sigma$.
\end{itemize}
    
As shown in Sec.~\ref{subsec:emrispectra} model \model{12} contains the highest number of unresolvable sources, which is even larger when the original catalog is processed through IFE.  
As our analysis does not reveal violations of stationarity and Gaussianity
, we expect our result to hold in the IFE-processed dataset.
The potential inclusion of the IFE algorithm would likely reinforce this finding, which has already been validated.

We additionally investigate the robustness of our findings on model \model{1}, relaxing the threshold for source removal described in Sec.~\ref{sec:gwb_eval} from $\rho_{\rm AK,th} = 1$ to $0.1$. 
Therefore, the number of sources effectively contributing to the GWB evaluation is increased from $N_{\rm final} = 26932$ to $209072$.
Fig. \ref{fig:gwb_A_1_01}, illustrates how the EMRI foreground for \model{1} changes when adding sources with starting approximated SNR in the range $0.1 < \rho_{\rm AK} < 1$. 
Non-coalescing EMRIs with $\rho < 1$ accumulate below $1 \rm mHz$, contributing to the low-frequency component of the GWB, with only a modest increase in SNR of about $2.5\%$, from ${\rm SNR}_{\rm gwb} = 311$ to $319$.
This is in line with our expectations:  
as coalescing binaries are typically resolvable with ${\rm SNR}=20$ up to $z=1$, EMRIs would have an SNR above $1$ should they be placed at the largest redshift in our integration range, $z<4.5$, with the exception of low-mass systems ($M\approx 10^4 M_\odot$) contributing only very little to the GWB. 
Conversely, a large number of non-coalescing EMRIs with $\rho < 1$ accumulates below 1mHz, hence the observed contribution GWB.
Moreover, low-frequency non-coalescing EMRIs are expected to be highly eccentric, which suppresses GW emission in the cross-polarization.

This aligns with findings from Ref.~\cite{2023PhRvD.108j3039P}, which shows a consistent $5\%$ reduction in background SNR after excluding the faintest sources across all examined models.

Fig.~\ref{fig:gwb_ray_M1} (right panel), shows that the Rayleigh test results remain largely unchanged. 
The addition of faint sources to the catalog does not substantially alter the foreground properties investigated.
In fact, including low SNR sources appears to primarily affect the signal PSD in Fig.~\ref{fig:gwb_A_1_01} only at $f<1 \rm mHz$, whereas the Rayleigh test in Fig.~\ref{fig:gwb_ray_M1} shows deviation at and above $f>1 \rm mHz$. 
Therefore, the choice of considering only sources with $\rho_{AK} > 1$ suffices to support our conclusions.

The threshold for detectability ($\rho = 20$), could similarly influence the GWB stationarity and its overall brightness. 
Establishing realistically EMRI detectability is heavily dependent on (i) an accurate detector characterization and (ii) a precise definition of the detection statistics. 
In absence of those, we opt to leave this for future work. 

\section{\label{sec:conclusions} Conclusions}

In this study, we conducted a statistical analysis of EMRI backgrounds as observed by LISA, focusing on the core assumptions of Gaussianity and stationarity. These are central in many aspects of LISA inference processes, and quantifying their validity is crucial. We constructed realistic EMRI foregrounds by combining state-of-the-art waveforms, a detailed description of the instrument response, and exploring a number of astrophysically-motivated populations. 
Using a test available in the literature, we analyzed the statistical properties of three representative populations, listed in Table~\ref{tab:models}.

Our results show varying degrees of violation, both in Gaussianity and stationarity. These are closely linked to the number of unresolvable sources contributing to each foreground: broadly speaking, brighter foregrounds are expected to yield smaller violations.
The implications of our findings are twofold. On one side, deviations exhibited by GWBs can be used to disentangle multiple overlapping ones.
In order to do so, a suitable Bayesian likelihood (beyond the Gaussian and stationary approximation) needs to be defined, and model selection performed with respect to the standard GWB spectral inference. Initial attempt at performing such inferences have recently appeared in literature~\cite{2023PhRvD.108j3005S,2024arXiv241008274P}. Our detection statistic offers a diagnosing tool (e.g. as a residual check) to inform the need to revise inference with extended models.
On the other side, and more importantly, PSD misestimates can introduce biases in resolvable source parameter reconstruction:  overestimating the PSD as a consequence of a Gaussian-likelihood model,  as shown in~\cite{2023PhRvD.108j3005S}, leads in the simplest scenario to overestimation of source luminosity distances. This could potentially result in a systematic offset of the source population reconstruction. 
More complex biases, induced by deviations with specific spectral shapes, may lead to even more complex systematics for intrinsic source parameters. 
In this context, our test can be used as a rapid flagging tool to quantifying the validity of the Gaussianity assumption during global fit execution. We leave a detailed assessment of the impact of such biases to future work.

To the best of our knowledge, this is the first statistical characterization of its kind applied to EMRI backgrounds. 
We envisage a number of possible future improvements.
The removal of resolvable sources should follow an iterative foreground estimation method, similarly to Ref.~\cite{2021PhRvD.104d3019K}.
Alternatively, and more robustly, one should consider directly employing full LISA global fit posteriors, obtained from simulated data including a population of EMRIs. Due to the lack of availability of such data, we were unable to pursue this approach in this work.

Moreover, a better understanding of the EMRI population will certainly provide more realistic predictions as compared to our approximations, e.g. to the CO population distribution. 
Finally, we point out that with the increasingly larger computational resources available, constructing EMRI foregrounds without pre-downselecting sources and including other astrophysical foregrounds in the analysis will likely become feasible.
Future work incorporating the suggested improvements will further refine our understanding of non-deterministic signals in the LISA datastream and contribute to the development of more robust data-analysis models for LISA.

\begin{acknowledgments}

The authors thank D.~Laghi, L.~Speri, N.~Tamanini, S.~Marsat, A.~Klein and V.~Gennari for useful insights and fruitful comments on this study.
M.P. and O.B. acknowledge support from the French space agency CNES in the framework of LISA.
F.P., M.B., A.S. R.B. acknowledges support through the Italian Space Agency grant \emph{Phase A activity for LISA mission, n. 2017--29--H.0}, by the MUR Grant ``Progetto Dipartimenti di Eccellenza 2023-2027'' (BiCoQ), and by the ICSC National Research Center funded by NextGenerationEU.\@
A.S. acknowledges financial support provided under the European Union’s H2020 ERC Consolidator Grant ``Binary Massive Black Hole Astrophysics'' (B Massive, Grant Agreement: 818691). M.B. acknowledges support provided by MUR under grant ``PNRR - Missione 4 Istruzione e Ricerca - Componente 2 Dalla Ricerca all'Impresa - Investimento 1.2 Finanziamento di progetti presentati da giovani ricercatori ID:SOE\_0163'' and by University of Milano-Bicocca under grant ``2022-NAZ-0482/B''.
Computational work was performed using at Bicocca's Akatsuki cluster (B Massive funded), CINECA with allocations through INFN, Bicocca, and ISCRA project HP10BEQ9JB.

\textit{Software}:
We acknowledge usage of the following additional
\textsc{Python} 
packages for modeling, analysis, post-processing, and production of results throughout:
\textsc{matplotlib}~\cite{2007CSE.....9...90H},
\textsc{numpy}~\cite{2020Natur.585..357H},
\textsc{scipy}~\cite{2020NatMe..17..261V}
\textsc{cupy}~\cite{cupy_learningsys2017}. This work makes use of the Black Hole Perturbation Toolkit~\cite{BHPToolkit}.
\end{acknowledgments}

\vfill
\bibliographystyle{apsrev4-2} 
\bibliography{main}
\end{document}